\documentclass[prl,twocolumn,aps,footinbib, preprintnumbers,amsmath]{revtex4-1}
\usepackage{color}
\usepackage[active]{srcltx}
\usepackage{amsmath,amsfonts,amssymb,amsthm,amstext,amscd,eucal,srcltx}
\usepackage{epsfig,graphicx,bm}
\usepackage{epstopdf, epsf}
\usepackage{dcolumn}
\usepackage{hyperref}
\newcommand{\ie}{{\em i.e. }}

\newcommand{\mpl}{M_{\rm pl}}
\newcommand{\be}{\begin{equation}}
\newcommand{\ee}{\end{equation}}

\newcommand{\bse}{\begin{subequations}}
\newcommand{\ese}{\end{subequations}}
\newcommand{\bea}{\begin{eqnarray}}
\newcommand{\eea}{\end{eqnarray}}
\newcommand{\ba}{\begin{array}}
\newcommand{\ea}{\end{array}}
\newcommand{\bc}{\begin{center}}
\newcommand{\ec}{\end{center}}

\begin{document}
\title{Gauge-flation: Inflation From Non-Abelian Gauge Fields }%
\author{A.~Maleknejad$^{1,2}$ and M.~M.~Sheikh-Jabbari$^2$}
\affiliation {$^1$Department of Physics, Alzahra  University
P. O. Box 19938, Tehran 91167, Iran,\\
$^2$School of Physics, Institute for research in fundamental
sciences (IPM), P.O.Box 19395-5531, Tehran, Iran }

\begin{abstract}
Inflationary models are usually based on dynamics of one or more
scalar fields coupled to gravity. In this work we present a new
class of inflationary models, \emph{gauge-flation} or non-Abelian
gauge field inflation, where slow-roll inflation is driven by a
non-Abelian gauge field. This class of models are based on a gauge
field theory with a generic non-Abelian gauge group minimally
coupled to gravity. We then focus on a particular gauge-flation
model by specifying the action for the gauge theory. This model has two parameters
 which can be determined using the current cosmological
data and has the prospect of being tested by Planck satellite data. Moreover, the values of these parameters are within the {natural} range of parameters
in generic grand unified theories of particle physics.

\end{abstract}
\maketitle


Inflationary Universe paradigm \cite{Inflation-Books, Weinberg},
the idea that early Universe has undergone an inflationary
(accelerated expansion) phase, has appeared very successful in
reproducing the current cosmological data through the $\Lambda$CDM
model \cite{Inflation-Books, Weinberg}. Many models of inflation
have been proposed and studied so far, e.g. see
\cite{Bassett-review}, which are all compatible with the current
data. Inflationary models are generically single or multi scalar
field theories with standard or non-standard kinetic terms and  a
potential term, which are minimally or non-minimally coupled to
gravity. Generically, in these models inflationary  period is driven
by a ``slowly rolling'' scalar field (inflaton field) whose  kinetic
energy  remains small compared to the potential terms.

Toward the end of inflation the kinetic term becomes comparable to
the potential energy, and inflaton field(s) start a (fast)
oscillation around the minimum of their potential losing their
energy to other fields present in the theory, the (p)reheating
period. The energy of the inflaton field(s) should eventually be
transferred to standard model particles, reheating, where standard
FRW cosmologies take over. Therefore, to have a successful cosmology
model one should embed the model into particle physics models. With
the current data the scale of inflation (or Hubble parameter $H$
during inflation) is not restricted well enough, it can range from
$10^{14}$ GeV to the Bing Bang Nucleosynthesis scale $1$ MeV.
However, larger $H$, $H\gtrsim 10$ GeV, is preferred within the
slow-roll inflationary models with preliminary particle or high
energy physics considerations. It is hence natural to tune the
inflationary model within the existing particle physics models
suitable for similar energy scales.

Most of successful inflationary scenarios so far use  scalar
field(s) as the inflaton, because turning on  time dependent scalar
fields does not spoil the homogeneity and isotropy of the cosmology.
Although it is relatively
easy to write down a potential respecting the slow-roll dynamics
conditions, it is generically not easy to argue for such potentials and their stability against quantum corrections within particle physics models. For example, the Higgs sector  in the ordinary
electroweak standard model minimally coupled to Einstein gravity does not   support a successful
inflationary model e.g. see \cite{Higgs-inflation}. The situation within
beyond standard model theories seems not to be better.

Vector gauge fields are commonplace in all particle physics models.
However, their naive usage in constructing inflationary models is in
clash with the homogeneity and isotropy of the background. It has
been argued that this obstacle may be overcome by introducing many
vector fields which contribute to the inflation, such that the
anisotropy induced by them all average out \cite{vector-inflation}. Alternatively one may introduce three
orthogonal vector fields and  retain rotational invariance by identifying each of these fields with
a specific direction in space \cite{vector-inflation}.
Nonetheless, it was shown that it is not possible to get a successful vector inflation model in a \emph{gauge invariant} setting \cite{vector-inflation}. Lack of gauge invariance, once  quantum
fluctuations are considered may lead to instability of the background and
may eventually invalidate the background classical inflationary dynamics
analysis \cite{vector-inflation-loophole}.

Here, we construct a new class of vector inflation models and to avoid the above mentioned possible instability issue we work in
the framework of gauge field theories. In addition, to remove the
 incompatibility with isotropy resulting from gauge fields
we introduce three gauge fields. We choose these gauge fields to
rotate among each other by $SU(2)$ non-Abelian \emph{gauge}
transformations. Explicitly, the rotational symmetry in 3d space is
retained because it is identified with the global part of the
$SU(2)$ gauge symmetry. In our model we need not restrict ourselves
to $SU(2)$ gauge theory and, since any non-Abelian gauge group has
an $SU(2)$ subgroup, our \emph{gauge-flation} (non-Abelian gauge
field inflation)  model can be embedded in non-Abelian gauge
theories with arbitrary gauge group. Another advantage of using
non-Abelian gauge theories is that, due to the structure of
non-Abelian gauge field strength, there is always a potential
induced for the combination of the gauge field components which
effectively plays the role of   the inflaton field.

In the above discussions we have only committed ourselves to the
gauge invariance and have not  fixed a specific gauge theory action.
This action will be fixed on the requirement of  having a successful
inflationary model. We study one such gauge-flation model but
gauge-flation models are expected not to be limited to this specific
choice. In this Letter we consider a simple two parameter
gauge-flation model and study classical inflationary trajectory for
this model as well as the cosmic perturbation theory around the
inflationary path. We then use the current data for constraining the
parameters of our model and show that our model  is compatible with
the current data within a natural range for its parameters.

\emph{\textbf{The inflationary setup.}}
Consider a 4-dimensional $su(2)$ gauge field $A^a_{~\mu}$, where
$a,b, ...$ and $\mu,\nu, ...$ are respectively used for the indices
of the gauge algebra and the space-time. We will be interested in
\emph{gauge invariant} Lagrangians ${\cal
L}(F^a_{~\mu\nu}, g_{\mu\nu})$ which are constructed out of metric
$g_{\mu\nu}$ and the strength field $F$%
\be\label{F-general}%
F^a_{~\mu\nu}=\partial_\mu A^a_{~\nu}-\partial_\nu
A^a_{~\mu}-g\epsilon^a_{~bc}A^b_{~\mu}A^c_{~\nu}, %
\ee%
where $\epsilon_{abc}$ is the totally antisymmetric tensor. We work
with FRW inflationary background metric%
\be \label{FRW} ds^2=-dt^2+a(t)^2\delta_{ij}dx^idx^j, %
\ee%
where indices $i,j, ...$ label the  spatial directions.

The effective inflaton field is introduced as follows:  We will work
in temporal gauge $A^a_{~0}=0$ and at the background level, as in
any inflationary model, we only allow for $t$ dependent field
configurations \cite{Galtsov}
\be\label{A-ansatz-background}
A^a_{~\mu}=\left\{
\begin{array}{ll} \phi(t)\delta^a_i\, ,\qquad  &\mu=i
\\
0\,, \qquad &\mu=0\,.
\end{array}\right.
\ee%
With this choice we are actually identifying our gauge indices with
the  spatial indices. That is, we identify the rotation group
$SO(3)$ with the global part of the gauge group, $SU(2)$. Therefore, the
rotational non-invariance resulted from turning on space components
of a vector is compensated by (the global part of) the gauge
symmetry.  $\phi(t)$ is not a genuine scalar, while%
\be\label{psi-def}%
\psi(t)=\frac{\phi(t)}{a(t)}%
\ee%
is indeed a scalar. (Note that for the flat FRW metric
$e^a_{~i}=a(t)\delta^a_i$ , where $e^a_{~i}$ are the 3d triads.) The
components of the field strengths in the ansatz  are
\be\label{F-components}%
F^a_{~0i}=\dot{\phi}\delta^a_i\,,\qquad
F^a_{~ij}=-g\phi^2\epsilon^a_{~ij}.
\ee%

After fixing the gauge and choosing $A^a_{~0}$ to be zero, system
has nine other degrees of freedom, $A^a_{~i}$. However, in the
ansatz \eqref{A-ansatz-background} we only keep one scalar degree of
freedom. We should hence first discuss consistency of the reduction
ansatz \eqref{A-ansatz-background} with the classical dynamics of
the system induced by
${\cal
L}(F^a_{~\mu\nu}, g_{\mu\nu})$. It is
straightforward to show that the gauge field equations of motion
$D_\mu\frac{\partial {\cal L}}{\partial F_{\mu\nu}}=0$,
where $D_\mu$ is the gauge covariant derivative, i) allows for a
solution of the form \eqref{A-ansatz-background} and, ii) once
evaluated on the ansatz \eqref{A-ansatz-background} becomes
equivalent to the equation of motion obtained from the ``reduced
Lagrangian'' ${\cal L}_{\textrm{red}}(\dot\phi,\phi; a(t))$,
\be\label{red-e.o.m}%
\frac{d}{a^3 dt}(a^3\frac{\partial {\cal L}_{red.}}{\partial
\dot\phi})-\frac{\partial {\cal L}_{red.}}{\partial \phi}=0 \,,
\ee%
where ${\cal L}_{red.}$ is obtained from inserting
\eqref{F-components} and metric \eqref{FRW} into the original gauge
theory Lagrangian ${\cal L}$. Moreover, one can show that the energy
momentum tensor, $T_{\mu\nu}$,
computed over the FRW background \eqref{FRW} and the gauge field
ansatz \eqref{A-ansatz-background} takes the form of a homogeneous
perfect fluid%
$$ T^{\mu}_{\ \nu}=diag(-\rho, P,P,P)\,,$$
which is the same as the energy momentum tensor obtained from the
reduced Lagrangian ${\cal L}_{\textrm{red}}$. That is,
\be\label{rho-P}%
\rho=\frac{\partial\mathcal{L}_{red.}}{\partial\dot{\phi}}\dot{\phi}-\mathcal{L}_{red.}\,,\qquad
P = \frac{\partial(a^3 {\cal L}_{red.})}{\partial a^3}\ .
\ee

All the above is true for any gauge invariant Lagrangian
${\cal L}={\cal L}(F_{\mu\nu}^a; g_{\mu\nu})$. To have a successful
inflationary model, however, we should now choose appropriate form of ${\cal
L}$. The first obvious choice is Yang-Mills action minimally coupled
to Einstein gravity. This will not lead to an inflating system with
$\rho+3P<0$, because as a result of scaling invariance of Yang-Mills
action one immediately obtains $P=\rho/3$ and that $\rho\geq 0$. So,
we need to consider modifications to Yang-Mills. As will become
clear momentarily one such appropriate choice
is%
\begin{equation}\label{The-model}%
S=\int
d^4x\sqrt{-{g}}\left[-\frac{R}{2}-\frac{1}{4}F^a_{~\mu\nu}F_a^{~\mu\nu}+\frac{\kappa}{384
}
(\epsilon^{\mu\nu\lambda\sigma}F^a_{~\mu\nu}F^a_{~\lambda\sigma})^2\right]\,
\end{equation}
where we have set $8\pi G\equiv \mpl^{-2}=1$ and
$\epsilon^{\mu\nu\lambda\sigma}$ is the totally antisymmetric
tensor. This specific $F^4$ term is chosen because the contribution
of this term to the energy momentum tensor will have the equation of
state $P=-\rho$, making it perfect for driving inflationary
dynamics. (To respect the weak energy condition for the $F^4$ term,
we choose $\kappa$ to be positive.) The reduced (effective)
Lagrangian is obtained from evaluating \eqref{The-model} for the
ansatz \eqref{A-ansatz-background}:
\be%
\mathcal{L}_{red}=\frac{3}{2}(\frac{\dot{\phi}^2}{a^2}-\frac{g^2\phi^4}{a^4}+\kappa
\frac{g^2\phi^4\dot{\phi}^2}{a^6}). %
\ee%
Energy density $\rho$ and pressure $P$ are then given by
\be\label{rho-P-total} %
\rho= \rho_{_{YM}}+\rho_\kappa\ ,\qquad P=\frac13\rho_{_{YM}}-\rho_\kappa, %
\ee%
where
\be\label{rho0-rho1}%
\rho_{_{YM}} =\frac{3}{2}(\frac{\dot{\phi}^2}{a^2}+\frac{g^2\phi^4}{a^4})\
, \qquad \rho_\kappa= \frac32\kappa \frac{g^2\phi^4\dot{\phi}^2}{a^6}\,.
\ee%
Recalling the Friedmann  equations%
\be\begin{split}
H^2&=\frac{1}{2}(\frac{\dot{\phi}^2}{a^2}+\frac{g^2\phi^4}{a^4}+\kappa
\frac{g^2\phi^4\dot{\phi}^2}{a^6}) ,\cr
\dot{H}~&=-(\frac{\dot{\phi}^2}{a^2}+\frac{g^2\phi^4}{a^4})\,,
\end{split}
\ee%
the slow-roll parameter $\epsilon$ is
\be\label{epsilon-slow-roll}%
\epsilon=-\frac{\dot H}{H^2}=\frac{2\rho_{_{YM}}}{\rho_{_{YM}}+\rho_\kappa}\,.
\ee%
To obtain a slow-roll inflationary phase initial conditions and
parameter $\kappa$ should be chosen such that $\rho_\kappa$ dominates
over $\rho_{_{YM}}$ during inflation. As slow-roll inflation progresses
the contribution of Yang-Mills term $\rho_{_{YM}}$ to the energy momentum
tensor grows and eventually at around $\rho_{_{YM}}=\rho_\kappa$ inflation
ends. To have a consistent slow-roll inflation, it is not sufficient
to have small $\epsilon$; for any physical quantity $X$, $\frac{\dot
X}{HX}$ should remain small. In particular, demanding the effective
scalar inflaton field $\psi$ to be slowly varying, \ie $\delta\equiv
-\frac{\dot\psi}{H\psi}\ll 1$
and $\dot\delta/(H\delta)\ll 1$,
yields 
\begin{subequations}
\begin{align}
\label{epsilon1}
\epsilon&\simeq\psi^2(1+\gamma),\quad \eta\simeq\psi^2\\
\label{delta-kappa}
\delta&\simeq\frac{\gamma}{6(\gamma+1)}\epsilon^2,\quad
\kappa\simeq\frac{(2-\epsilon)(1+\gamma)^3}{g^2\epsilon^3},
\end{align}%
\end{subequations}
 in the
leading order in $\epsilon$. In the above%
\be\label{x-def}%
 \gamma=\frac{g^2\psi^2}{H^2}\quad \textrm{or
equivalently}\quad
H^2\simeq\frac{g^2\epsilon}{\gamma(\gamma+1)}\,,%
\ee%
$\gamma$ is a slowly varying positive parameter of order one.
Since $\delta\sim \epsilon^2$ (\emph{cf.} \eqref{delta-kappa}), $\psi$ is varying slower than $\epsilon$
and hence from \eqref{x-def} we learn that during slow-roll regime%
\be
\frac{\epsilon}{\epsilon_i}\simeq\frac{\gamma+1}{\gamma_i+1}\,,\quad \frac{\gamma}{\gamma_i}\simeq\frac{H_i^2}{H^2}\,,
\ee%
where $\epsilon_i,\ \gamma_i$ and $H_i$ are the values of these parameters at
the beginning of inflation. Number of e-folds $N_e$ at the end of
 inflation, marked by $\epsilon_f=1$, is
then given by%
\bea \label{Ne}%
N_e=\int_{t_i}^{t_f} Hdt=-\int_{H_i}^{H_f} \frac{dH}{\epsilon
H}\simeq \frac{\gamma_i+1}{2\epsilon_i}\ln \frac{\gamma_i+1}{\gamma_i}\,. %
\eea%
The value of $\psi$ at the beginning and end of inflation are related as $\psi_f^6\simeq \frac{1}{2}\psi_i^6$, where \eqref{delta-kappa} has been used and by $\simeq$ sign we mean equality to the leading order in slow-roll
parameter $\epsilon$. Notice that all the dimensionful quantities,
like $\kappa,\ \psi$ and $H$, are measured in units of $\mpl$.

\emph{\textbf{Gauge-flation cosmic perturbation theory.}}
Sofar we have analyzed  dynamics of the homogeneous effective scalar
inflaton field $\psi$,  while consistently turning off the other
gauge field components. To compare our model with the data we should
 work out the power spectrum of curvature perturbations and their
spectral tilt for which we need to study cosmic perturbation
theory in gauge-flation. In general small fluctuation around the
ansatz \eqref{A-ansatz-background} can be parameterized by 12 fields
$\delta A_\mu^a$. Decomposing $\mu$ index into time and spatial
parts and identifying the gauge index $a$ with the spatial index
$i$, these 12 fields give rise to four scalars, three
divergence-free vectors and a divergence-free, traceless
symmetric tensor:%
\bea \label{gauge-field-fluct-decomp}%
\delta
A^a_{~0}&=&\delta^{ak}\partial_k\dot{Y}+\delta^a_j u^j,\nonumber\\
\label{a2}
\delta A^a_{~i}&=&\delta^a_i Q+\delta^{aj}\partial_{ij}(M+\partial_i v_j+t_{ij})+\epsilon^{a~j}_{~i}(g\phi\partial_{j}P+w_j), \nonumber\eea %
where $\partial_i$ denotes partial derivative respect to $x^i$, the
scalars are parameterized by $Y, Q, M, P$, vectors by $u_i, v_i,\
w_i$ and the tensor by $t_{ij}$. As we see, we are indeed dealing
with a multi-field inflationary model. Among the
scalars, $Q$ can be identified with the fluctuation of the inflaton
field $\phi$.

The other field active during inflation is metric
whose fluctuations are customarily parameterized
by four scalars, two divergence-free vectors and one tensor:%
\bea
&&ds^2=-(1+2A)dt^2+2a(\partial_iB-S_i)dx^idt\nonumber\\
&+&a^2\left((1-2C)\delta_{ij}+2\partial_{ij}E+2\partial_{(i}W_{j)}+h_{ij}\right)dx^idx^j.\,\,\,\nonumber
\eea%
In the first order perturbation theory which we are interested in,
scalar, vector and tensor fluctuations do not couple to each
other.
Among  12 gauge field perturbations and 10 metric perturbations one scalar and one vector mode of the gauge field, and two scalars and one vector of the metric modes are gauge degrees of freedom. We
hence remain with five gauge-invariant scalar, three massless vector and two massless tensor modes.

\begin{figure*}
 \centerline{\psfig{file=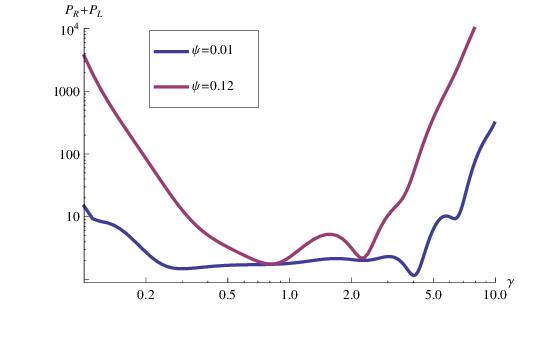, width=3.3in} \psfig{file=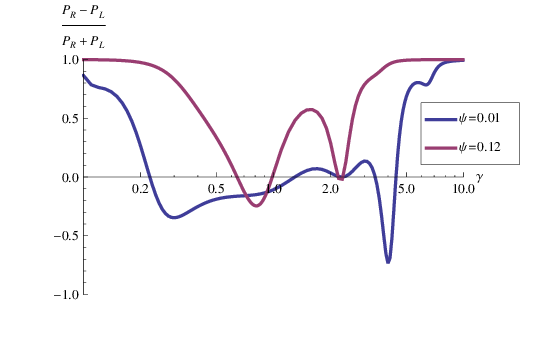, width=3.3in}}
\caption{In the left panel we have depicted ${P_{_{R}}}+{P_{_{L}}}$. In the standard scalar-driven inflationary models ${P_{_{R}}}={P_{_{L}}}=1$. The right panel the parity violating factor $\frac{P_{_{R}}-P_{_{L}}}{P_{_{R}}+P_{_{L}}}$ versus $\gamma$ for $\psi=10^{-2}$ and $\psi=0.12$ is shown. The power spectra have been calculated at $k\tau=-0.01$, long enough after modes have crossed the horizon and behave quite classically. As we see in the right panel,  for very small and very large $\gamma$ values $P_{_{R}}\gg P_{_{L}}$.}\label{Tensor-Mode-power-figures}
\end{figure*}

Equations of motion for the perturbations can be obtained from
perturbed Einstein equations $\delta G_{\mu\nu}=\delta T_{\mu\nu},$
which decomposes into four equations for scalar modes, two for vector modes and one equation for tensor modes
\cite{Weinberg}. The equation of motion for the remaining  scalar, vector and tensor mode $t_{ij}$  is provided through  perturbed gauge field equations.

A thorough analysis reveals that amplitude of vector perturbations are exponentially
suppressed, as in the ordinary scalar-driven
inflationary models \cite{gauge-flation-prd}.
Although the tensor mode perturbations in the gauge field sector $t_{ij}$ are suppressed at the superhorizon scales, their presence leads to \textit{parity violating} terms in the second order action governing the metric tensor perturbations  $h_{ij}$ \cite{gauge-flation-prd}. This happens due to the fact that one of two modes of $t_{ij}$ (say, the right-handed circular polarization)  just before the horizon-crossing undergoes a tachynoic growth for a short period and as a result  the right-handed circular polarization of $h_{ij}$ becomes large at superhorizon value. On the other hand, the left-handed polarization of $t_{ij}$ remains small at  horizon-crossing and has negligible effect on the superhorizon value of its corresponding $h_{ij}$ polarization.

The power spectra for the Left and Right gravitational wave modes are obtained as \cite{gauge-flation-prd}
\bea\label{PT}%
\mathcal{P}_{T_{R}}\simeq  P_{_R} \left(\frac{H}{\pi}\right)^2\!\!\big\vert_{k=aH}\quad\&\quad
\mathcal{P}_{T_{L}}\simeq  P_{_L} \left(\frac{H}{\pi}\right)^2\!\!\big\vert_{k=aH}\,,\nonumber%
\eea%
where $P_{_R}$ and $P_{_L}$ are functions of the parameters $\gamma,\psi$ (Fig.\ref{Tensor-Mode-power-figures}). The power spectrum of the tensor modes, is then given as $\mathcal{P}_{T}=\mathcal{P}_{T_{R}}+\mathcal{P}_{T_{L}}$.

The full analysis of cosmic perturbation theory in our model has
many new and novel features compared to the standard scalar-driven
inflationary models, a
detailed analysis of which is presented in \cite{gauge-flation-prd}, in the following table we summarize the results:\\
\begin{center}
\begin{tabular}{|p{4cm} | p{1cm}| p{2cm}|}
\hline
 \emph{Power spectrum of   curvature perturbations}& $\mathcal{P}_{\mathcal{R}}$  & $\frac{1}{8\pi^2\epsilon}\left(\frac{H}{\mpl}\right)\!^2$  \\ [2.5ex]
\hline
 \hspace{1cm} \emph{Spectral Tilt}&  $n_s-1$ & $\hspace{0.2cm} -2(\epsilon-\eta)$  \\ [2.5ex]
\hline
 \hspace{0.25cm} \emph{Tensor to Scalar ratio}& \hspace{0.3cm}$r$ & $8(P_{_{R}}+P_{_{L}})\epsilon$ \\ [2.5ex]
\hline
\hspace{0.25cm} \emph{Power spectrum of  anisotropic inertia $a^2\!\pi^S$ }& $\mathcal{P}_{\!a^2\!\pi^S}$ &  $\frac{\epsilon}{8\pi^2}\left(\frac{H}{\mpl}\right)\!^2$ \\ [2.5ex]
\hline
\end{tabular}\label{gauge-flation-summary-table}\end{center}
A  specific feature of gauge-flation is that it predicts a non-zero power spectrum for the scalar anisotropic inertia  $a^2\!\pi^S$ \cite{Weinberg}, with the ratio
\be
\frac{\mathcal{P}_{\!a^2\!\pi^S}}{\mathcal{P}_{\mathcal{R}}}=\epsilon^2\,.
\ee
Note that $a^2\!\pi^S$ is identically zero in all the scalar-driven inflationary models in the context of Einstein GR.

\emph{\textbf{Confronting gauge-flation with the data.}} To this end, we depict the results of our model on the  allowed region of the $n_s-r$ graph:

\begin{figure*}
\centerline{\psfig{file=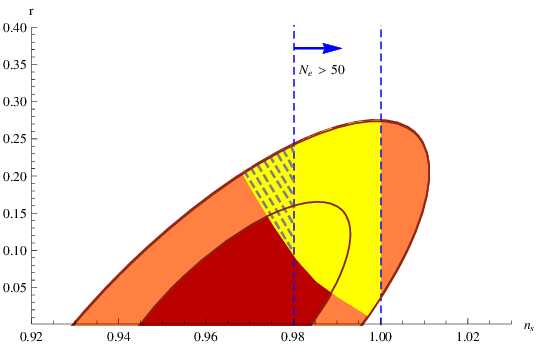, width=3in} \psfig{file=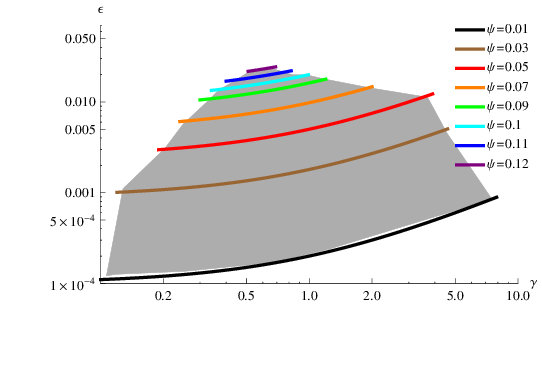, width=3in}}
\caption{The left panel shows  $1\sigma$ and $2\sigma$ contour bounds of 7-year WMAP+BAO+H0. The yellow area (region with lighter color) represents the gauge-flation predictions for $\psi\in(0.01,0.12)$ range. The region with enough number of e-folds restricts us to $n_s>0.98$ region, that is on the right-side of the $N_e=50$ line \cite{gauge-flation-prd}. Therefore, the allowed region is the highlighted region between $N_e=50$ and $n_s=1$ lines. The shaded region in right panel shows the allowed values for $\epsilon$ and $\psi$. }\label{ns-r-gf}
\end{figure*}
From the left panel of Figs \ref{ns-r-gf}, we learn that in the allowed region the value of $\gamma$ is restricted as
$\gamma\in(0.1-8)$  which determines the value of $\kappa$ and $g$
\be\begin{split}
g &\simeq(0.15 - 3.7)\times 10^{-3},\cr 
\Lambda &\sim (10^{-5}-10^{-4})\mpl\,,\quad \kappa\equiv \Lambda^{-4}\,.
\end{split}\ee
Restricting ourselves to $1\sigma$ contour in Fig \ref{ns-r-gf}, we find stringent bounds on $r$, $n_{\mathcal{R}}$, $\frac{\mathcal{P}_{\!a^2\!\pi^S}}{\mathcal{P}_{\mathcal{R}}}$ and $H$
\bea\label{2sigma-values}
0.05<&r&<0.15\ ,\ \quad H\simeq (3.4-5.4)\times 10^{-5}\mpl,\\
0.98 \leq & n_s&\leq 0.99\ ,\ \quad \frac{\mathcal{P}_{\!a^2\!\pi^S}}{\mathcal{P}_{\mathcal{R}}}\simeq (3.6-22)\times 10^{-5},
\eea
while within the $1\sigma$ contour, we have $0.5<\gamma<4$
and the gauge field value during inflation turns out to be sub-Planckian  $\psi\simeq (0.4-1)\times 10^{-1}\mpl$.

\emph{\textbf{Discussion.}} We showed that non-Abelian gauge field driven inflation, \emph{gauge-flation}, can lead to a successful slow-roll inflation model with specific features. In the model we considered the theory has two parameters, gauge coupling $g$ and the coefficient of the $(F\tilde F)^2$ term $\kappa$.
The value for the gauge coupling $g$ required by the CMB data is of order $10^{-3}$, while $\Lambda$, the scale associated with $\kappa\sim\Lambda^{-4}$, is of order $10^{14}$GeV. These two parameters are in the natural range for perturbative beyond standard models of particle physics. Moreover, the $\kappa$-term may be obtained by integrating out axionic fields where $\Lambda$ is associated with scale of the axion potential \cite{Weinberg-vol2,{integrate-out}}. For this procedure to be theoretically meaningful we need $\Lambda\gg H$, which is respected by the best-fit values of our model.

Current data tightly restricts the values of our parameters. In particular, noting Fig. \ref{ns-r-gf}, our model {predicts} that the tensor-to-scalar ratio $r$ is restricted to be in $0.02<r<0.15$ range, which is well within the
range to be probed by the Planck satellite. As another prediction, while gauge-flation has always a red spectral tilt, the tilt has a lower bound $n_s>0.98$.

Finally we point out a specific  feature of our model not shared by usual scalar-driven inflationary models: gauge-flation predicts a non-zero scalar anisotropic inertia $(a^2\!\pi^S\neq0)$, and  $\frac{\mathcal{P}_{a\!^2\!\pi\!^S}}{\cal{P}_{\cal{R}}}\sim10^{-4}$. It would be interesting to explore observational prospects this ratio, which we postpone to future works.

\textbf{Note added:} More than a year after appearance of the original version of this work on the arXiv, the paper \cite{AMW} appeared which prompted us to recheck and correct the tensor mode sector of gauge-flation cosmic perturbation theory. A more detailed analysis may be found in \cite{{gauge-flation-prd},{The-review}}.

\textbf{Acknowledgement}: We would like to thank Amjad Ashoorioon and Bruce Bassett  for
comments on the manuscript and Niayesh Afshordi for useful
discussions and comments. Work of A.M. is partially supported by
grants from Bonyad-e Melli Nokhbegan of Iran.




\end{document}